\documentstyle[11pt,newpasp,twoside,psfig]{article}
\def\edcomment#1{\iffalse\marginpar{\raggedright\sl#1\/}\else\relax\fi}
\reversemarginpar

\begin{document}
\newcommand{\sauron}{{\tt SAURON}}

\title{Probing the stellar populations of early-type galaxies: \\
	the \sauron\ survey}
 \author{Eric Emsellem}
\affil{$^1$ Centre de Recherche Astronomique de Lyon, Saint Genis Laval, France}
\author{Roger Davies, Richard McDermid}
\affil{$^2$ Physics Departement, University of Durham, Durham, UK}
\author{Harald Kuntschner}
\affil{$^3$ European Southern Observatory, Garching bei M\"unchen, Germany}
\author{Reynier Peletier}
\affil{$^4$ Department of Physics and Astronomy, University of Nottingham, Nottingham, UK}
\author{Roland Bacon$^1$, Martin Bureau$^6$, Michele Cappellari$^5$, Yannick Copin$^7$, Bryan Miller$^8$,
Ellen Verolme$^5$, Tim de Zeeuw$^5$}
\affil{$^5$ Sterrewacht Leiden, Leiden, The Netherlands} 
\affil{$^6$ Department of Astronomy, Columbia University, New York, USA}
\affil{$^7$ Institut de Physique Nucl\'eaire de Lyon, Villeurbanne, France}
\affil{$^8$ Gemini Observatory, La Serena, Chile}

\begin{abstract}
The \sauron\ project will deliver two-dimensional spectroscopic data
of a sample of nearby early-type galaxies with unprecedented quality.
In this paper, we focus on the mapping of their stellar populations
using the \sauron\ data, and present some preliminary results on
a few prototypical cases.
\end{abstract}

\section{Introduction}

In recent years very significant progress has been made in 
both the theoretical and observational study of the stellar populations
of early-type galaxies. Stellar population synthesis models now rely
on detailed stellar evolution tracks and much more complete stellar libraries
at higher resolution (e.g. Vazdekis et al. 1996, Kodama \& Arimoto 1997). 
Recent work on nearby samples of galaxies have often focused on the 
breaking of the age-metallicity degeneracy, following the widely used Lick/IDS 
indices to probe dynamically hot galaxies (J{\o}rgensen 1999).
Kuntschner (2000) emphasized the effect of non-solar abundance particularly
when interpreting the observed line indices of early-type, metal rich, galaxies:
Fornax ellipticals thus exhibit a relatively narrow spread in age, the
distribution of galaxies in index/index diagrams being mostly attributed to 
differences in metallicity.

Although the conclusions reached by these studies do indeed converge towards
a scenario consistent with the so-called hierarchical universe, 
the question originally posed by the Lick group two decades ago
is still pertinent: when and how do early-type galaxies form? 
Most observational campaigns along these lines have considered line-strength gradients
only, and where thus only able to probe a one-dimensional (long-slit) section
of the galaxies.
Two-dimensional spectroscopy has been mostly used for probing the kinematics
of galaxies. Except in a handful of cases (e.g. Emsellem et al. 1996, Peletier et al. 1999,
del Burgo et al. 2001), the available stellar absorption lines have only served
a surprisingly restricted purpose.

One of the aims of the \sauron\ project is to provide the 2D
spatial coverage which is missing. In this paper, we describe
the approach the \sauron\ group is following, and present
some preliminary results regarding the stellar populations for
a few prototypical examples of galaxies in the \sauron\ sample.

\section{From where, to where?}

The essential ingredients of a good recipe for the formation and evolution
of early-type galaxies should include a treatment of the star
formation history, together with a scheme that links this to their
dynamics and morphology. We first sum up briefly what we
know about their chemical and dynamical properties.

\subsection{What do we know?}

Early-type galaxies are found preferentially in high local density regions.
Their properties vary, from giant luminous ellipticals which are assumed
to be triaxial, pressure supported objects with shallow nuclear brightness profiles,
to low luminosity ellipticals, which exhibit disky isophotes, are consistent
with isotropic rotators and include steep power law central cusps.
In this context, the bulges of early-type spirals resemble low-luminosity ellipticals.
Most if not all galaxies are thought to harbour central supermassive black holes,
the masses of which correlate with the mean stellar dispersion of the spheroid
(Ferrarese \& Merritt 2000, Gebhardt et al. 2000).

Colours of ellipticals become redder with higher luminosity, an effect usually attributed
to a parallel increase in metal enrichment and alpha element abundance.
The star formation timescale in ellipticals is thought to be short,
and ellipticals in general are viewed as old, coeval systems in which, as already
emphasized, metallicity increases with luminosity. S0s have a larger 
spread in (luminosity weighted) age than ellipticals (Kuntschner \& Davies 1998,
Kuntschner 2000; see Fig.~1).
\begin{figure}
\centerline{\psfig{figure=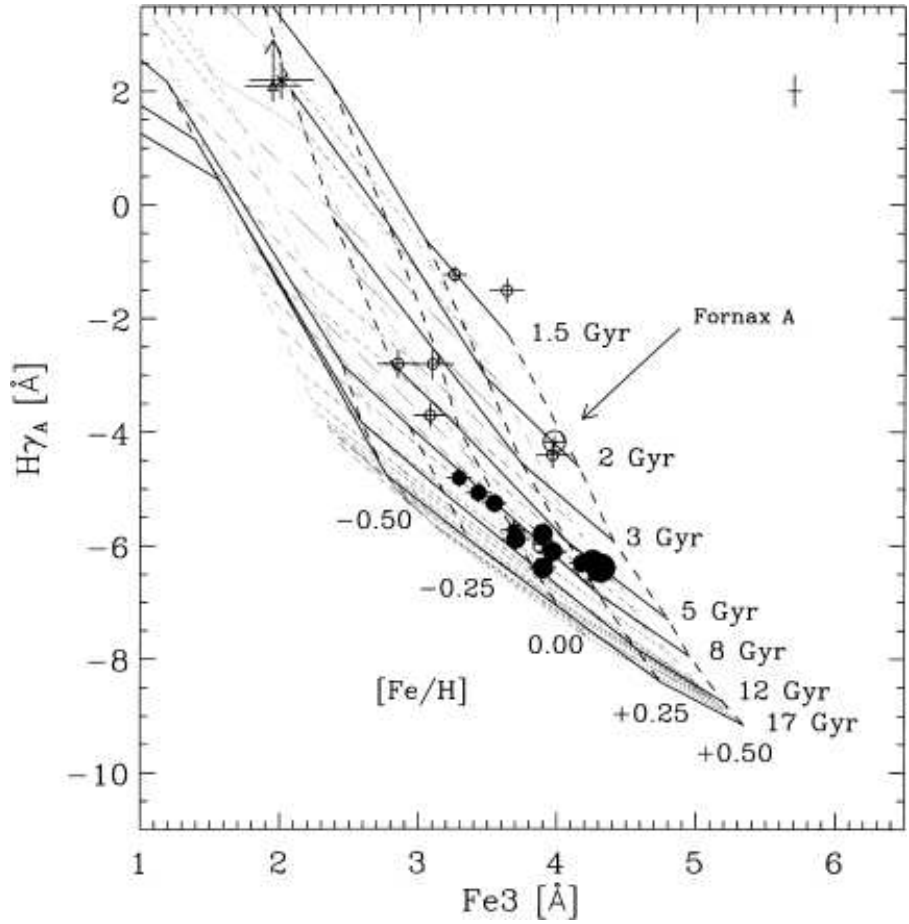,width=12cm}}
\caption{H$\gamma$ versus Fe3 diagram. Filled circles and open circles represent ellipticals
and S0s respectively. The stars and open triangles represent possible post-starburst and starburst galaxies
respectively. The cross in the upper right corner indicates the rms uncertainty in the 
transformation to the Lick/IDS system. From Kuntschner 2000.}
\end{figure}

As we look back in time, star formation rates increase and interactions and mergers become
more common. Already by $z=0.5$, the fraction of blue galaxies in rich clusters
has increased dramatically compared to $z=0$, 
a trend known as the Butcher Oemler effect (Butcher \& Oemler 1984).
Furthermore HST studies (Dressler et al. 1997) have shown that
these blue galaxies are spirals that are 2 to 3 more common in intermediate redshift
clusters with lenticular galaxies being correspondingly less common.
The signatures of mergers are observed in todays lenticulars and ellipticals 
as e.g. shells, kiloparsec scale decoupled cores, or counter-rotating gas disks.

\subsection{What would we like to know?}

Any scenario for the formation and evolution of early-type galaxies 
should explain the basic changes in their properties in terms of
luminosity, Hubble type and environment.
What information can we hope to dig out?
Galaxies (besides our own) are observed projected on the sky,
leaving us two spatial dimensions, and only one (line-of-sight) velocity.

The first obstacle is therefore to solve the inverse problem of 
determining the intrinsic shapes of these galaxies.
In dynamical terms, we can push things further by determining
the dominant orbit families which constitute their skeleton.
These certainly hold traces of the involved formation and evolution processes,
including passive secular evolution. Then questions, more directly related to
the hierarchical merging scenario, follow: how common are decoupled cores, how common
and massive are black holes in galactic nuclei?

Other integrated quantities such as colours and line indices can then
help us to examine the stellar population content of these galaxies.
How common are the disks in ellipticals, and is their stellar population
different (younger?) from the rest of the galaxy. Can we constrain the 
stellar populations of the decoupled cores, and more generally
of individual orbit families? 
To determine the origin of the different morphological/dynamical components
we therefore need to reconcile the kinematics and stellar populations.

\subsection{SAURON: the goals}

The \sauron\ project was motivated by the questions summarized above. 
The minimal requirements 
included the two-dimensional mapping of the kinematics and line strengths
(including at least one age and one metallicity indicator) of a reasonable
sample of early-type galaxies up to at least one half $R_e$. 
We first designed a dedicated integral field spectrograph, to cover
a field of view of 33$\times$41~arcsec$^2$, and to work
in the spectral domain around the Mg triplet at 5170~\AA\ that
includes Mg and Fe and H$\beta$ features as well as the [OIII] emission lines
(see Bacon et al. 2001 for a description of the instrument).
We then defined a representative sample of 72 galaxies, 24 for each
E, S0 and Sa classes, evenly distributed in cluster and field type environments
(de Zeeuw et al. 2002).

The \sauron\ observational campaign, representing more than 50 nights
at the William Herschel Telescope (La Palma), is producing
data with unprecedented uniformity and quality, 
and represents a unique database to unravel the stellar population - kinematics
connection.
	
\section{Age and composition of a galaxy}

In practice, how do we determine the stellar population content of a galaxy,
when it cannot be resolved into individual stars?
The first step, adopted for stellar synthesis modeling, is to simulate
the spectrum of a cluster whose stars share a single age and metallicity.
Such a single burst spectrum can be built assuming an initial mass function (IMF),
and following the stars along their evolution in the HR diagram.
Each stellar spectrum is then matched using corresponding observed spectra.
The final composite spectrum is the sum of the individual spectra 
weighted according to their respective contributions.
This can be repeated for a grid of different ages and metallicities,
and even different IMFs, to provide a set of template synthesized spectra.
The observed galaxy spectrum can then be fitted, providing
values for the luminosity-weighted mean age and metallicity.

Apart from the fact that a galaxy spectrum includes the contributions
of stars with different ages and metallicities (integrated along the
line of sight), the fitting of the full spectrum is sometimes not optimal
as it may amplify errors in the flux calibration, does not
necessarily focus on the relevant information, and requires
very high signal-to-noise data. Since the age and metal enrichment are
directly reflected in the depths of the absorption lines, these can be
used as tracers of the stellar populations.

\subsection{The Lick system}

The Lick/IDS system of indices was thus designed with the purpose of 
focusing on important absorption features. These indices are easily measured,
and can be compared with the values obtained from stellar synthesis models.
Some indices are well known to be more age sensitive (e.g. hydrogen
Balmer lines), or metalliticity sensitive (e.g. Fe lines).

However the Lick/IDS system was defined about 2 decades ago, and has
a few limitations:
\begin{itemize}
\item The spectral resolution used for the Lick system is $\sim 8.4$~\AA,
diluting the available spectral signatures.
\item The standard stellar spectra used to calibrate the Lick system 
were not flux calibrated.
\item As defined, the Lick indices are rather sensitive to the 
required correction for velocity broadening.
\item The measurements of the Lick stars are based on a non-linear
detector.
\end{itemize}
These factors arise largely from the increase in efficiency and
precision of modern instruments. Although we now see their limitations
the Lick indices have played a pivotal role in the study of early-type
galaxy populations for almost 30 years.

\subsection{New models, indices and libraries}

However, recent stellar synthesis models, such as those
developed by Vazdekis and collaborators (Vazdekis 1999), do not just
provide line indices, but also predict the spectrum over the \sauron\ range
at a spectral resolution of 1.8~\AA! This gives us the ability to
define new, better adapted, indices, and to operate thorough spectral
fitting of the galaxy data, even allowing for mixtures of different stellar populations.

In the context of the \sauron\ survey, we have therefore designed a new index, 
based on the original Fe5270 Lick one: it is less dependent on 
the broadening correction, and more sensitive to metallicity.
New age-metallicity grids have been derived using the Vazdekis models which
include up-to-date isochrones and stellar libraries.
We have also observed a significant number of stars with \sauron, including
high-metallicity ones, in order to compile a consistent library of
stellar templates (Bacon et al. 2001).

\section{Line strength maps with SAURON}

\subsection{Validation of the instrument}

The first step before interpreting the \sauron\ line-strength maps
is to validate the data itself by checking whether our line-strength 
measurements agree with previously published values.
We have done this for a number of galaxies for which line-strengths are available
in the literature, and Fig.~2 shows such a comparison in the case of NGC~5813.
Overall, the agreement is excellent up to the edge of the \sauron\ field
of view. There are some slight offsets between the different data sets, but these
are within the quoted error bars.
\begin{figure}[h]
\centerline{\psfig{figure=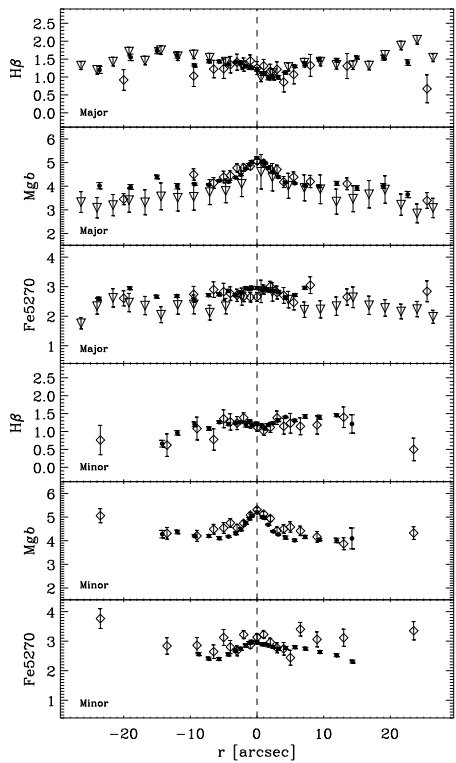,width=6cm}}
\caption{Comparison between \sauron\ and published values of the line-strengths
in NGC~5813. The three upper (resp. lower) panels are taken along the major
(resp. minor) axis. \sauron\ values are indicated as black dots, diamonds are from
Gorgas et al. 1990, MNRAS 215, 217, and triangles are from Gonzalez 1993 (PhD Thesis).}
\end{figure}

\subsection{Emission lines and how to get rid of them}
\begin{figure}
\centerline{\psfig{figure=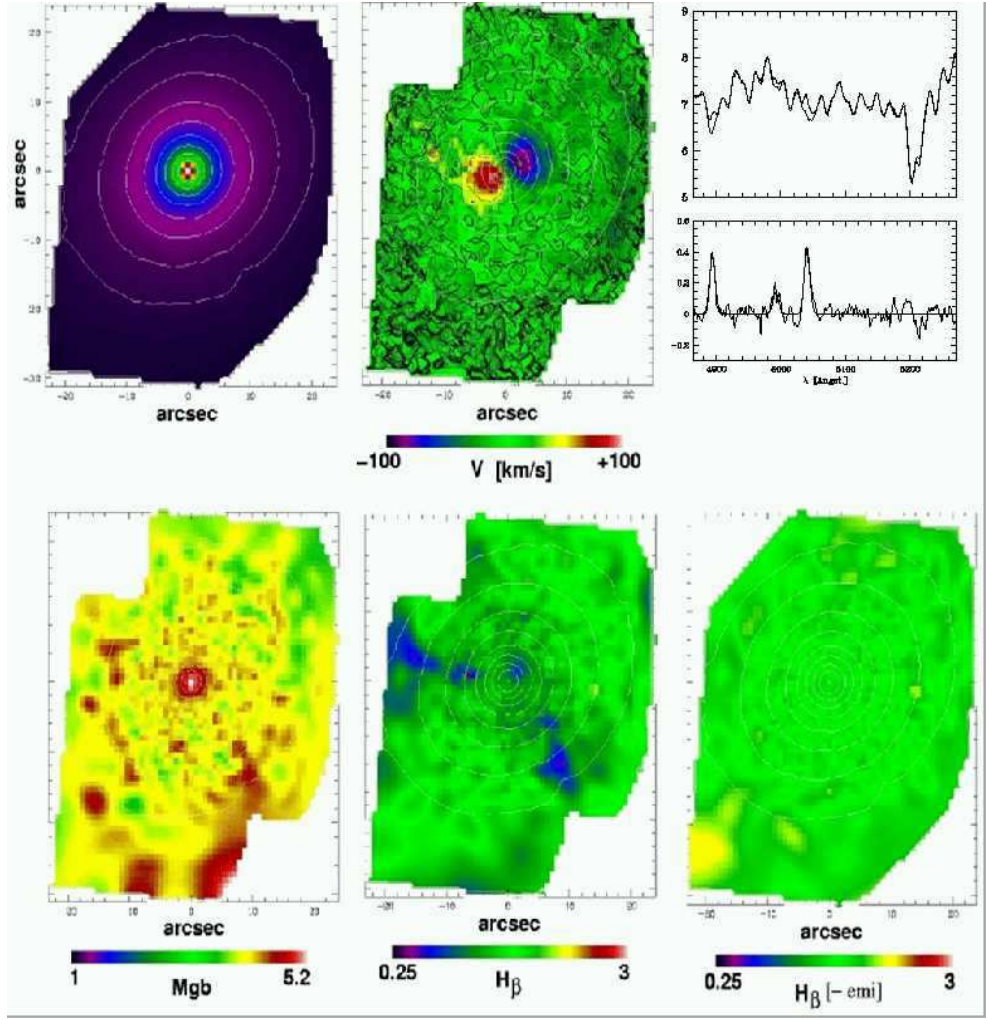,width=13cm}}
\caption{\sauron\ kinematics and line-strengths for NGC~5813.
Top panels: the reconstructed image (surface brightness, left) and
stellar velocity field (middle). The top right panel shows a fit of the stellar
contribution of one central spectrum, and the residual emission lines fitted
by three Gaussians corresponding to the $H\beta$, and [OIII] lines.
Bottom panels: \sauron\ maps for Mgb, H$\beta$ before and H$\beta$ after correction
for the emission lines contributions.}
\end{figure}

One important issue when deriving line-strengths is to make sure that
emission lines are not contaminating the spectra, as this can
sometimes be a major source of error particularly for the H$\beta$ and 
Mgb indices (Goudfrooij \& Emsellem 1996). In order to correct for the
contribution of emission lines, we have designed a simple three-step procedure.
We first fit the spectra using a complete spectral library which includes
both stars and galaxies without nebular emission. This fit is achieved
by masking the spectral regions where emission lines may be present.
The fit obviously requires the knowledge of the stellar kinematics 
(velocity and dispersion), also obtained by avoiding the contaminated regions.
We then subtract the fits to obtain pure emission line spectra, which
are themselves fitted using different components for each emission line
(H$\beta$, [OIII], [NI]). These emission-line fits are finally subtracted 
from the original spectra, (ideally) leading to emission-line free spectra. 
Line-strengths are measured from the emission-corrected spectra. 

This procedure is illustrated in the case of NGC~5813, the central region of which
exhibits the presence of emission line gas. Fig.~3 shows the H$\beta$ map before
and after correction for the emission line contributions. In the former,
there is a clear structure of lower H$\beta$ values which corresponds
to a gaseous filament.
As shown in the top right panel of Fig.~2, the H$\beta$ and [OIII]$\lambda$4959
emission lines are barely visible in the original spectra, but
appear clearly in the residual pure emission-line spectra.
After correction, the H$\beta$ map is flat over the full \sauron\ field
of view, and shows no evidence for the presence of the
decoupled core which is apparent only in the stellar velocity field.

\section{SAURON results: two decoupled cores}

In this Section, we present preliminary results of the \sauron\
observations of two decoupled cores, to illustrate the importance
of two-dimensional spectroscopy in probing the stellar populations
of early-type galaxies.
\begin{figure}
\centerline{\psfig{figure=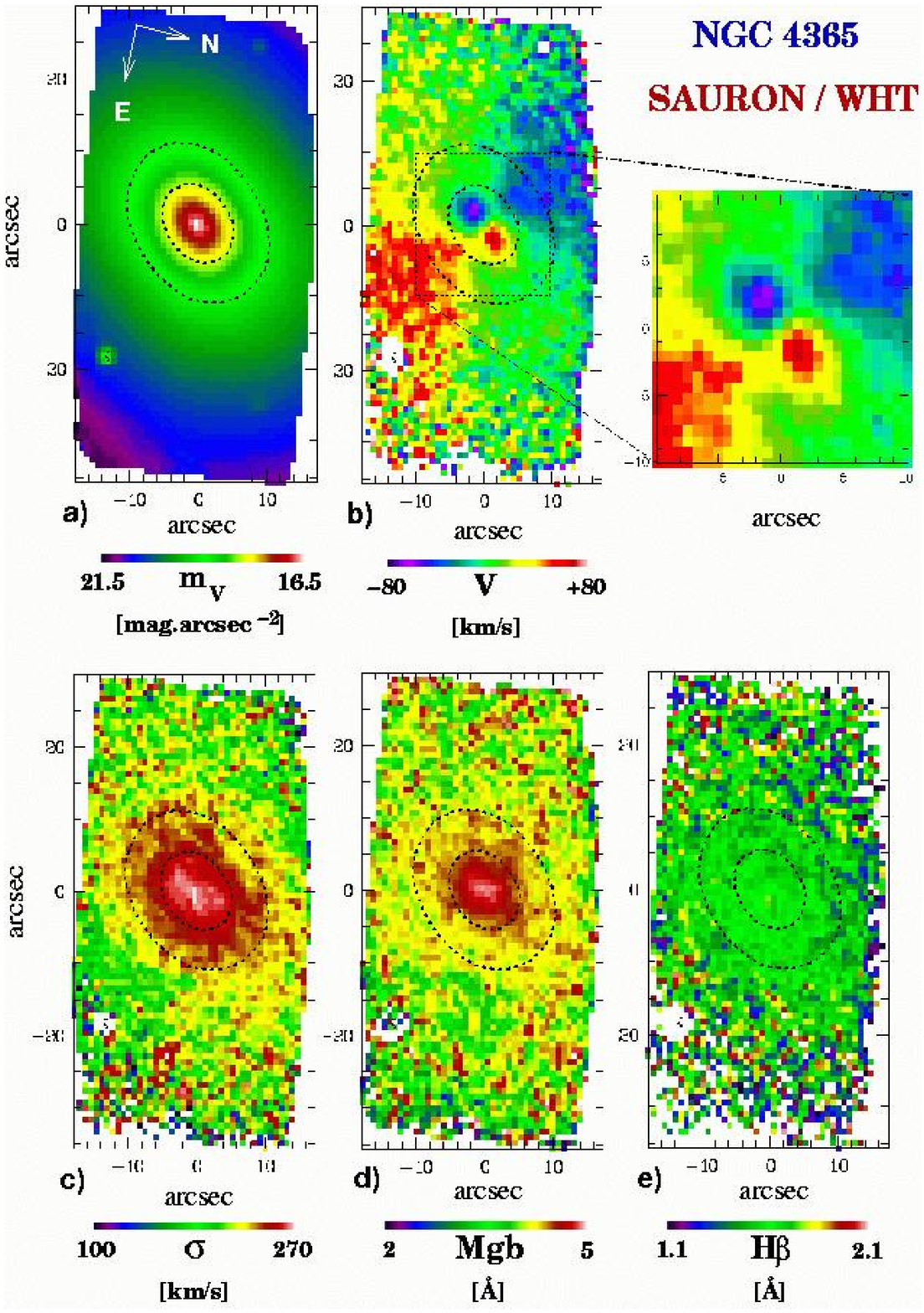,width=12cm}}
\caption{\sauron\ maps of NGC4365. From top to bottom, left right:
reconstructed image, stellar velocity, zoomed stellar velocity,
dispersion, Mgb and H$\beta$.}
\end{figure}

\subsection{NGC 4365: the classical case}

NGC~4365 is an elliptical known to harbour a kinematically decoupled
core (Surma \& Bender 1995). The kinematic major-axis of the core was assumed 
to lie at 90 degrees from that of the outer prolate-like body. However the \sauron\
kinematical maps revealed a more complex morphology with the zero velocity
curve having a slowly varying position angle in the outer part (Davies et al. 2001).
This strongly hints at a triaxial geometry the details of which can
only be constrained by the full two-dimensional coverage provided by \sauron.

The line-strength maps lead to more surprising results. The Mgb increases
towards the centre, with isocontours which closely resemble the galaxy isophotes.
The H$\beta$ map is flat over the whole field, except perhaps for a slight
unresolved peak at the very centre of the galaxy (Fig.~4). The stellar population
of the core and the main body are found to be indistinguishable
in age, metallicity and abundance ratios (Fig.~5). This strongly suggests
that the decoupled core of NGC~4365 is the result of an early accretion,
the core and the outer body of the galaxy being roughly coeval. It also
tells us that the structure observed in NGC~4365 corresponds to a stable dynamical
configuration.

\subsection{NGC 4150: the post-starburst}

The case of NGC~4150, an S0 galaxy, is at first sight even more puzzling.
The stellar velocity field shows the presence of a small counter-rotating
core with a low velocity amplitude of about 20 km/s peak to peak.
The Mgb map of NGC~4150 shows a central dip, and an enhanced H$\beta$
in the central few arcseconds! When projected on a H$\beta$ versus
[MgFe5270] model grid, this corresponds to a slightly decreasing metallicity
and strongly decreasing age towards the centre (Fig.~6). It thus seems that
the central region of NGC~4150 was the scene of a recent gaseous accretion event
followed by a burst of star formation.
\begin{figure}[h]
\centerline{\psfig{figure=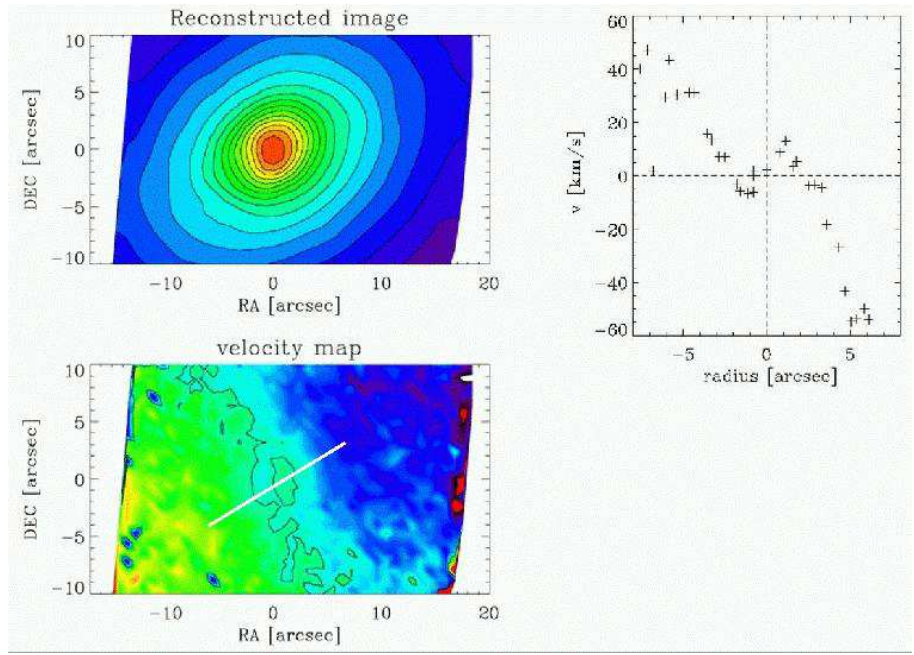,width=12cm}}
\caption{\sauron\ observations of the S0 galaxy NGC~4150.
Left panels: \sauron\ reconstructed intensity (top) and stellar velocity 
(bottom) maps. Right panel: velocity cut along the slit as drawn on the
stellar velocity field. The counter-rotating core is clearly apparent here.}
\end{figure}

\begin{figure}
\centerline{\psfig{figure=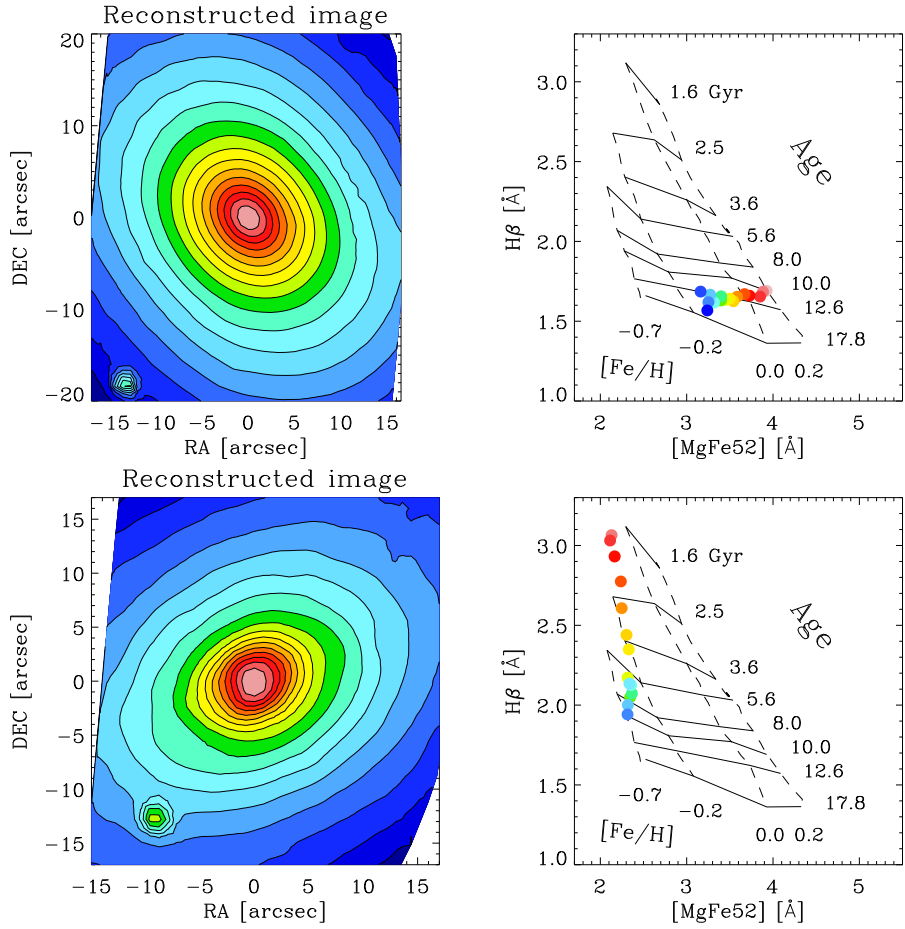,width=12cm}}
\caption{\sauron\ reconstructed maps (left panels)
and H$\beta$ versus [MgFe5270] indices (right panels)
for 2 kinematically decoupled cores: NGC~4365 (top) and NGC~4150 (bottom).
The core and main body of NGC~4365 have nearly indistinguishable ages,
metallicity and abundance ratios, and the core of NGC~4150 is very probably
the remnant of a recent gaseous accretion followed by star formation.}
\end{figure}

\section{Conclusions}

In this paper, we presented the cases of NGC~4365 and NGC~4150 to illustrate
the importance of two-dimensional spectroscopy for studying the stellar
populations of galaxies. However, the complexity and richness in the structures
of early-type galaxies will only be fully revealed when the complete set 
of \sauron\ datacubes has been analysed. 

The \sauron\ campaign will end in April '02. We will then have
a unique data set to probe the stellar content of early-type galaxies.
We are also gathering new stellar libraries and state-of-the-art stellar 
synthesis models. As emphasized in this paper, this is a critical step 
if we wish to fully exploit the \sauron\ datacubes. The efforts of 
the \sauron\ team are therefore not only
directed to acquiring the observations for the representative
sample of 72 early-type galaxies, but are also focused on the development
of new analysis tools and models. These tools will allow us to 
link the chemical and dynamical history of ellipticals and early-type
spirals, and to ultimately constrain the formation and evolution scenarios.

\end{document}